\documentclass[prb,showpacs,twocolumn]{revtex4}
\usepackage{graphicx}
\usepackage{dcolumn}
\usepackage{bm}
\usepackage{amsmath}

\makeatletter
\def\ExtendSymbol#1#2#3#4#5{\ext@arrow
6095{\arrowfill@#1#2#3}{#4}{#5}}
\newcommand\myequal[2][]{\ExtendSymbol{=}{=}{=}{#1}{#2}}
\makeatother

\begin{document}

\title{Transport properties for a Luttinger liquid wire with Rashba
spin-orbit coupling and Zeeman splitting}

\author{Fang Cheng$^{1,2}$}
\author{Guanghui Zhou$^{1,2,3}$}
\email{ghzhou@hunnu.edu.cn}

\affiliation{$^1$CCAST (World Laboratory), PO Box 8730, Beijing
             100080, China}

\affiliation{$^2$Department of Physics, Hunan Normal University,
             Changsha 410081, China\footnote{Mailing address}}

\affiliation{$^3$International Center for Materials Physics,
             Chinese Academy of Sciences, Shenyang 110015, China}

\begin{abstract}
We study the transport properties for a Luttinger-liquid (LL)
quantum wire in the presence of both Rashba spin-orbit coupling
(SOC) and a weak external in-plane magnetic field. The bosonized
Hamiltonian of the system with an externally applied longitudinal
electric field is established. And then the equations of motion for
the bosonic phase fields are solved in the Fourier space, with which
the both charge and spin conductivities for the system are
calculated analytically based on the linear response theory.
Generally, the ac conductivity is an oscillation function of the
strengths of electron-electron interaction, Rashba SOC and magnetic
field, as well as the driving frequency and the measurement position
in the wire. Through analysis with some examples it is demonstrated
that the modification on the conductivity due to electron-electron
interactions is more remarkable than that due to SOC, while the
effects of SOC and Zeeman splitting on the conductivity are very
similar. The spin-polarized conductivities for the system in the
absence of Zeeman effect or SOC are also discussed, respectively.
The ratio of the spin-polarized conductivities
$\sigma_\uparrow/\sigma_\downarrow$ is dependent of the
electron-electron interactions for the system without SOC, while it
is independent of the electron-electron interactions for the system
without Zeeman splitting.

\end{abstract}

 \vspace{0.2cm}

\maketitle

\section{INTRODUCTION}
The physics of one dimensional (1D) systems of strongly correlated
particles has become a very interesting subject because of the
simplicity of the models and the attainment of the truly 1D systems
due to breakthrough in material technology. From the theoretical
point of view, Luttinger-liquid (LL) model is appropriate to
describe the transport properties of 1D systems with
electron-electron interactions.$^1$ The LL model is of fundamental
importance because it is one of very few strongly correlated ``non
Fermi liquid" systems that can be analyzed in any detail. The model
does not attempt a complete description of electrons in a 1D metal
but rather confines to the vicinity of the Fermi surface. One of the
key features of the LL model is the spin-charge separation: the
low-energy excitations are not quasiparticles with charge $e$ and
spin $\hbar/2$ together but collective modes of spin and charge
excitation separately. Therefore, quantum transport in LL systems
has attracted a great deal of interest since the experimental
realization of the narrow quantum wire formed in semiconductor
heterostructures$^2$ and the carbon nanotube$^3$ as well as the edge
states of the fractional Quantum Hall liquid.$^4$ We will use the
first one as our physical subject in this work.

Spintronics is a multidisciplinary field whose central theme is the
active manipulation of spin degrees of freedom in solid state
systems. It is believed to be a promising candidate for the future
information technology.$^5$ There are two physical mechanisms which
can be used to influence the dynamics of the electron spin in normal
conductors, i.e., spin-orbit coupling (SOC) and Zeeman splitting. In
layered semiconductors devices, the two predominant types of SOC are
Dresselhaus SOC$^{6}$ and Rashba SOC.$^{7}$ The former arises from
the breaking of inversion symmetry by the inherent asymmetry of the
atomic arrangement in the structure and is not very amenable to
external manipulation. The latter, on the other hand, arises from
band bending at the interfaces between semiconductor layers and/or
any external electric fields applied to the device. Unlike
Dresselhaus SOC, the strength of the Rashba SOC can be partially
controlled by application of an external electric field
perpendicular to the two-dimensional electron gas (2DEG)
plane.$^{8}$ And in many of the proposed spintronics device
structures the spin manipulation relies on the Rashba SOC, and as
such, only the Rashba SOC will be considered in our work.

In recent few years, there have been tremendous published research
works of the SOC effects on the III-V type and II-VI type
nonmagnetic semiconductor heterostructures for the purpose of
spintronics devices. But there have been only few works$^{9-15}$
concerning electron-electron interactions in these spintronic
systems. The early theoretical studies$^{9}$ demonstrated that the
influence of the Zeeman splitting for a LL quantum wire is the
breaking of spin-charge separation, where the ratio of the spin-up
and spin-down conductivities in a dirty system diverges at low
temperatures due to the electron correlation and results in a
spin-polarized current. Further studies$^{10-12}$ also have shown
that the effect of Rashba SOC for the LL wire is that the spin
degeneracy is lifted for $k\neq0$ and each branch loses its vertical
symmetry axis, i.e., different directions of motion have different
Fermi velocities. Moreover, Coulomb corrections to extrinsic spin
Hall-effect of a 2DEG has also been studied recently.$^{13}$ In
methodology, a bosonization theory including a Rashba SOC$^{10}$ or
a Zeeman splitting$^9$ has been constructed. A further question
arising naturally is what will happen if both the SOC and Zeeman
splitting are considered simultaneously. This motivation has led to
the studies$^{14,15}$ on the combined presence of a Rashba SOC and a
Zeeman effect in an interacting quantum wire. In these works, the
study on a Coulomb long-ranged electron interaction quantum
wire$^{15}$ in the combined presence of a Rashba SOC and a Zeeman
splitting using the perturbative renormalization group treatment has
indicated the generation of a spin pseudogap and the propagation of
a well-defined spin-oriented current, and on a LL quantum
wire$^{14}$ by bosonization technique has demonstrated that the
tunneling current may deviate from a simple power law which is that
in an ordinary LL wire.

On the other hand, however, the pure LL quantum wire without SOC and
Zeeman splitting has been extensively investigated in last decade.
The ac response of 1D interacting system has been studied$^{16-18}$
previously in the framework of the LL model with or without
impurity. Moreover, it is known that the dc limit of conductance
through a clean LL quantum wire is not renormalized by the
electron-electron interactions when the reservoirs (leads) are
properly taken into account because a well-known phenomenon of
Andreev-type reflection occurring at the contact between a LL
quantum wire and a non-interacting reservoir.$^{19}$

In this work, we study the ac dynamical transport properties for a
homogenous interacting quantum wire in the presence of both internal
Rashba SOC and external magnetic field simultaneously. When a
longitudinal time-varying electric field is applied to the wire, in
the LL regime we use the bosonization technique and solve the
equation of motion for the system in the Fourier space. It is found
that in this case the spin and charge degrees of freedom are
completely coupled and can be characterized by four new different
velocities. Within the linear response theory, the dynamical ac
$(\omega\neq0)$ conductivity of the system is generally an
oscillation function of the strengths of electron-electron
interaction, Rashba SOC and Zeeman interaction as well as the
driving frequency and the measurement position in the wire. However,
the modification of $g$ on the conductivity is more remarkable than
the modification on the conductivity due to electron-electron
interactions is more remarkable than that due to SOC, while the
effects of SOC and Zeeman splitting on the conductivity are very
similar. The spin-polarized conductivities for the system in the
absence of Zeeman effect or SOC are also discussed, respectively.
The ratio of the spin-polarized conductivities
$\sigma_\uparrow/\sigma_\downarrow$ is dependent of the
electron-electron interactions for the system without SOC, and the
ratio becomes more different when the electron-electron interactions
are stronger, while it is independent of the electron-electron
interactions for the system without Zeeman splitting. To the best of
our knowledge, some of these phenomena have not been reported
previously for the LL quantum wire system.

The rest of the paper is organized as follows. In Sec. II, we
formulate the model Hamiltonian in the bosonization form for an
interacting quantum wire simultaneously with an external
longitudinal electric field applied and both Rashba SOC and Zeeman
splitting, and solve the equation of motion for the bosonic phase
fields in the Fourier space. Within the linear response theory, the
conductivity of the system is analytically calculated in Sec. III,
and the detailed results for the two limited cases without either
Zeeman splitting or Rashba SOC are demonstrated in two subsections,
respectively. Some examples and the discussion of the results are
demonstrated in Sec. IV. Finally, Sec. V concludes the paper.

\section{The Hamiltonian and Bosonization}
Consider the system consisting of an interacting 1D quantum wire
with an applied longitudinal electric field realized by an
externally applied electromagnetic radiation in the experience. In
the 1D quantum wire the electron is subjected to a Rashba SOC. Here
we have taken the symmetric center of the quantum wire as the origin
and the growth direction of the heterostructure to be the $z$ axis
in our spatial coordinate system. The electron transport
ballistically in the quantum wire along the longitudinal $x$
direction. A magnetic field $B$ perpendicular to the quantum wire is
applied along $y$ axis.

For a weak magnetic field, its coupling to the electron orbital can
be neglected$^{14}$ if the low-lying excitation is considered, so we
only keep the Zeeman Hamiltonian term with respect to the magnetic
field. Assuming $\delta v_R\sim\delta v_B\ll v_F$, the linearized
noninteracting electron Hamiltonian of the quantum wire with both
Rashba SOC and Zeeman splitting is given by$^{9,10,14}$
\begin{equation}
\label{myeq1} H_0=-i\hbar\int\sum_{\gamma,s}v_\gamma^s\psi_{\gamma
s}^{+}\partial_{x}\psi_{\gamma s}dx.
\end{equation}
The operators $\psi_{\gamma s}$ $(\gamma=L,R;s=\downarrow,\uparrow)$
annihilate  spin-down $(\downarrow)$ and spin-up $(\uparrow)$
electrons near the left (L) and right $(R)$ Fermi points. In what
follows, the indices $\gamma$ and $s$ take the values -1 (1) for
$L(R)$ and $\downarrow(\uparrow)$, respectively. And
$v_\gamma^s=\gamma v_F-\frac{1}{2}s\delta v_R+\frac{1}{2}\gamma
s\delta v_B$ are four different sound velocities. Here $v_F$ is the
bare Fermi velocity of noninteracting right and left movers, $\delta
v_R=2\alpha/\hbar$ ($\alpha$ is the strength of Rashba SOC) and
$\delta v_B=g'\mu_{B}B/k_{F}$ ($B$ is the magnitude of magnetic
field, $g'$ is the Lande factor, and $\mu_{B}$ is the Bohr magneton,
respectively). Eq. (\ref{myeq1}) shows clearly that the Rashba term
splits horizontally the bands and makes the electron Fermi
velocities become different for different directions of motion,
while the Zeeman term splits vertically the bands and makes the
electron Fermi velocities become different for different directions
of spin. Using the bosonization technique$^{20}$ in terms of
\begin{equation}
\label{myeq2} \psi_{\gamma s}^{+}\partial_{x}\psi_{\gamma
s}=i\gamma\bigg(\frac{\gamma\partial_x\vartheta_s-
\frac{\displaystyle \Pi_s}{\displaystyle \hbar}}{2}\bigg)^2,
\end{equation}
we can derive Hamiltonian (1) as
\begin{eqnarray}
\label{myeq3} H_0&=&\frac{\hbar v_F}{2}\int
dx\Big[(\partial_x\vartheta_{\uparrow})^2+(\frac{\Pi_{\uparrow}}
{\hbar})^2+(\partial_x\vartheta_{\downarrow})^2+
(\frac{\Pi_{\downarrow}}{\hbar})^2\Big]\nonumber\\
&-&\frac{\hbar}{2}\delta v_B\int
dx\Big[(\partial_x\vartheta_{\downarrow})^2+(\frac{\Pi_{\downarrow}}{\hbar})^2
-(\partial_x\vartheta_{\uparrow})^2-(\frac{\Pi_{\uparrow}}{\hbar})^2\Big]\nonumber\\
&+&\frac{\delta v_R}{2}\int
dx\Big[\Pi_{\uparrow}(\partial_x\vartheta_{\uparrow})-\Pi_{\downarrow}
(\partial_x\vartheta_{\downarrow})\Big],
\end{eqnarray}
where $\vartheta_{\uparrow/\downarrow}$ is the phase field for
spin-up/down electrons and $\Pi_{\uparrow/\downarrow}$ is the
corresponding conjugate momentum. With the transformation
\begin{equation}
\label{myeq4}
\vartheta_\rho=\frac{\vartheta_\uparrow+\vartheta_\downarrow}{\sqrt{2}},~~
\vartheta_\sigma=\frac{\vartheta_\uparrow-\vartheta_\downarrow}{\sqrt{2}},
\Pi_\rho=\frac{\Pi_\uparrow+\Pi_\downarrow}{\sqrt{2}},~~
\Pi_\sigma=\frac{\Pi_\uparrow-\Pi_\downarrow}{\sqrt{2}},
\end{equation}
we can reduce Eq. (\ref{myeq3}) into
\begin{eqnarray}
\label{myeq5} H_0&=&\frac{\hbar v_F}{2}\int
dx\Big[(\partial_x\vartheta_\rho)^2+ (\frac{\Pi_\rho}{\hbar})^2+
(\partial_x\vartheta_\sigma)^2+(\frac{\Pi_\sigma}{\hbar})^2\Big]
\nonumber\\
&+&\frac{\hbar}{2}\delta v_B\int dx\Big[(\partial_x\vartheta_\sigma)
(\partial_x\vartheta_\rho)+\frac{1}{\hbar^2} \Pi_\rho\Pi_\sigma\Big]\nonumber\\
&+&\frac{\delta v_R}{2}\int dx\Big[\Pi_\sigma
(\partial_x\vartheta_\rho)+\Pi_\rho
(\partial_x\vartheta_\sigma)\Big],
\end{eqnarray}
where $\vartheta_{\rho}$ and $\vartheta_{\sigma}$ can be considered
as the phase field corresponding to the charge degree and the spin
degree of freedom, respectively, and $\Pi_\rho$ and $\Pi_\sigma$ are
the corresponding conjugate momentums.

Next, the short-ranged electron-electron interactions in the wire
give a term to the Hamiltonian
\begin{equation}
\label{myeq6} H_{int}=\frac{V(q=0)}{2\pi}\int
dx(\partial_x\vartheta_\rho)^2,
\end{equation}
where $V(q=0)$ is the electron-electron interaction potential. In
this Hamiltonian we have neglected the Umklapp scattering, which is
not relevant in the quantum wires formed in semiconductor
heterostructure.

Finally, we consider the Hamiltonian term of a longitudinal electric
field applied to the quantum wire. We use the method of describing
the application of an external bias voltage.$^{19}$ With the
electron charge $-e$, the coupling to an external time-dependent
potential $U_R(t)$ yields a term in the Hamiltonian as$^{17}$
\begin{equation}
\label{myeq7} H_{ac}=-e\int
dx\rho(x)U_R(x,t)=\sqrt{\frac{2}{\pi}}e\int dx\partial_x
U_R(x,t)\vartheta_\rho,
\end{equation}
where $U_R(x,t)$ is the chemical potential of the right moving
electrons,
$\rho(x,t)=\sqrt{\frac2{\pi}}\partial_x\vartheta_{\rho}(x,t)$ is the
charge density in bosonization presentation. By virtue of the
relation $\partial_xU_R(x,t)=-E(x,t)$, Eq. (\ref{myeq7}) can be
expressed as
\begin{equation}
\label{myeq8} H_{ac}=-\sqrt{\frac{2}{\pi}}e\int
dxE(x,t)\vartheta_\rho(x,t),
\end{equation}
where $E(x,t)$ is the externally applied electric field.

Combining Eqs. (\ref{myeq5}), (\ref{myeq6}) and (\ref{myeq8}), we
finally obtain the total bosonized Hamiltonian for the system
\begin{eqnarray}
\label{myeq9} H&=&\frac{\hbar}{2}\int
dx\Big[\frac{v_\rho}{g}(\partial_x\vartheta_\rho)^2+v_F
(\frac{\Pi_\rho}{\hbar})^2\Big]\nonumber\\&+&\frac{\hbar}{2}\int
dx\Big[v_\sigma(\partial_x\vartheta_\sigma)^2+v_\sigma
(\frac{\Pi_\sigma}{\hbar})^2\Big]\nonumber\\&+&
\frac{\hbar}{2}\delta v_B\int dx\Big[(\partial_x\vartheta_\sigma)
(\partial_x\vartheta_\rho)+\frac{1}{\hbar^2}
\Pi_\rho\Pi_\sigma)\Big] \nonumber\\&+&\frac{\hbar}{2}\delta v_R\int
dx\Big[(\frac{\Pi_\sigma}{\hbar})(\partial_x\vartheta_\rho)+(\frac{\Pi_\rho}{\hbar})
(\partial_x\vartheta_\sigma)\Big] \nonumber
\\&-&\sqrt{\frac{2}{\pi}}e\int dx E(x,t)\vartheta_\rho(x,t),
\end{eqnarray}
where $v_{\rho,\sigma}$ are the propagation velocities of the charge
and spin collective modes of the decoupled model ($\delta v_B=\delta
v_R=0$, or $B=\alpha=0$) and the parameter $g$ is the strength of
the electron-electron interactions, which is defined as
$1/g^2=1+V(q=0)/\hbar\pi v_F$ with $v_F$ is the non-interacting
Fermion velocity of the system, non-interacting Fermions corresponds
to $g=1$ and repulsive interaction corresponds to $g<1$. The
velocities $v_{\rho,\sigma}$ have been obtained as function of $g$
and $v_F$ in Ref.[20] as $v_\rho=v_F/g$ and $v_\sigma=v_F$ for the
decoupled model.

Further, the action functional of the coupled system can be written
in terms of the phase fields $\vartheta_\rho(x,t)$ and
$\vartheta_\sigma(x,t)$ as
\begin{eqnarray}
\label{myeq10} S&=&\frac{\hbar}{2}\int dt\int
dx\Big[\frac{1}{gv_\rho}(\partial_t\vartheta_\rho)^2-\frac{v_\rho}{g}
(\partial_x\vartheta_\rho)^2\Big]\nonumber\\&+&\frac{\hbar}{2}\int
dt\int dx\Big[\frac{1}{v_\sigma}
(\partial_t\vartheta_\sigma)^2-v_\sigma
(\partial_x\vartheta_\sigma)^2\Big]\nonumber\\&-&\frac{\hbar}{2}\frac{\delta
v_R}{v_F}\int dt\int dx\Big[(\partial_t\vartheta_\sigma)
(\partial_x\vartheta_\rho)+(\partial_t\vartheta_\rho)
(\partial_x\vartheta_\sigma)\Big]\nonumber\\&-&
\frac{\hbar}{2}\delta v_B\int dx\Big[(\partial_x\vartheta_\sigma)
(\partial_x\vartheta_\rho)+\frac{1}{(v_F)^2}(\partial_t\vartheta_\rho)
(\partial_t\vartheta_\sigma)\Big]\nonumber
\\&+&\sqrt{\frac{2}{\pi}}e\int dx E(x,t)\vartheta_\rho(x,t).
\end{eqnarray}
Note that in our system the time derivative of the field is not
anymore proportional to the conjugate canonical momentum, but is a
linear combination of the canonical momentum (including charge
canonical momenta and spin canonical momenta) and the gradient of
the field. However, after omitting the second power of the shifts
from the spin-obit or the Zeeman term and the product between them,
we find that the extra charge or spin canonical momentum and the
gradient of the field produce the same terms in the first line of
Eq. (9) and the first term
$(\Pi_\rho\partial_t\vartheta_\rho+\Pi_\sigma\partial_t\vartheta_\sigma)$
of the Lagrangian $L=\int dx
(\Pi_\rho\partial_t\vartheta_\rho+\Pi_\sigma\partial_t\vartheta_\sigma)-H$,
which finally make that the extra charge or spin canonical momentum
and the gradient of the field have not effect on the action
functional of the coupled system. Therefore, by minimizing action
(\ref{myeq10}) we obtain its associated equations of motion for the
phase fields
\begin{widetext}
\begin{align}
\label{myeq11}
\frac{\hbar}{gv_\rho}\partial_{t}^2\vartheta_{\rho}-\frac{\hbar
v_\rho}{g}\partial_{x}^2\vartheta_{\rho} -\hbar\frac{\delta
v_R}{v_F}\partial_{t}\partial_{x}\vartheta_{\sigma}
-\frac{\hbar}{2}\delta v_B\partial_{x}^2\vartheta_{\sigma}
-\frac{\hbar}{2}\frac{\delta
v_B}{{v_{F}}^2}\partial_{t}^2\vartheta_{\sigma}+\sqrt{\frac{2}{\pi}}eE(x,t)=0,\\
\label{myeq12}
\frac{\hbar}{v_\sigma}\partial_{t}^2\vartheta_{\sigma}-\hbar
v_\sigma\partial_{x}^2\vartheta_{\sigma} -\hbar\frac{\delta
v_R}{v_F}\partial_{t}\partial_{x}\vartheta_{\rho}
-\frac{\hbar}{2}\delta v_B\partial_{x}^2\vartheta_{\rho}
-\frac{\hbar}{2}\frac{\delta
v_B}{{v_{F}}^2}\partial_{t}^2\vartheta_{\rho}=0.
\end{align}
\end{widetext}
Applying the Fourier transformation
\begin{equation}
\label{myeq13} \vartheta(x,t)=\frac{1}{(2\pi)^2}\int dq \int d\omega
\vartheta(q,\omega)e^{-iqx+i\omega t}
\end{equation}
to Eqs. (\ref{myeq11}) and (\ref{myeq12}), we have the solution for
the phase fields
\begin{widetext}
\begin{align}
\label{myeq14} \vartheta_{\rho}(q,\omega)=\sqrt{\frac{2}{\pi}}
\frac{ev_{F}}{\hbar} \frac{E(q,\omega)} {\displaystyle
(\omega^2-v_{\rho}^{2}q^2)- \frac{\displaystyle (\delta
v_Rq\omega-\frac{1}{2} \delta v_Bv_Fq^2-\frac{1}{2} \frac{\delta
v_B}{v_F}\omega^2)^{2}}{\displaystyle
\omega^{2}-v_{\sigma}^{2}q^2}},\\
\label{myeq15} \vartheta_{\sigma}(q,\omega)=-\sqrt{\frac{2}{\pi}}
\frac{ev_{F}}{\hbar} \frac{(\delta v_Rq\omega-\frac{1}{2} \delta
v_Bv_Fq^2-\frac{1}{2} \frac{\delta v_B}{v_F}\omega^2)E(q,\omega)}
{\displaystyle
(\omega^2-v_{\rho}^{2}q^2)(\omega^{2}-v_{\sigma}^{2}q^2)-(\delta
v_Rq\omega-\frac{1}{2}\delta v_Bv_Fq^2-\frac{1}{2} \frac{\delta
v_B}{v_F}\omega^2)^{2}}.
\end{align}
\end{widetext}

\section{Conductivity of the system}
The current operator can be defined by using the 1D continuity
equation $\partial_xj_{\rho}(x,t)=e\partial_t\rho(x,t)$. Then we
have the charge current
\begin{equation}
\label{myeq16}
j_{\rho}(x,t)=\sqrt{\frac{2}{\pi}}e\partial_t\vartheta_{\rho}(x,t).
\end{equation}
Therefore, using solution (\ref{myeq14}) for
$\vartheta_\rho(q,\omega)$, we obtain the explicit expression for
the charge current operator
\begin{widetext}
\begin{equation}
\label{myeq17}
j_{\rho}(q,\omega)=\frac{ie^2v_F}{\hbar\pi}\frac{2\omega(\omega^{2}-v_{\sigma}^{2}q^2)
E(q,\omega)}{(\omega^2-v_{\rho}^{2}q^2)(\omega^{2}-v_{\sigma}^{2}q^2)-(\delta
v_{R}q\omega-\frac{1}{2} \delta v_{B} v_{F}q^2-\frac{1}{2}
\frac{\delta v_{B}}{v_{F}}\omega^2)^2},
\end{equation}

And the charge current operator is written further as
\begin{equation}
\label{myeq18} j_{\rho}(q,\omega)=\frac{ie^2v_F}{\hbar\pi}\frac{i
E(q,\omega)}{1-\frac{\displaystyle \delta v_B^2}{\displaystyle
4v_{F}^{2}}}
\frac{2\omega(\omega^2-v_{\sigma}^{2}q^2)}{\displaystyle
(\omega+u_1q)(\omega+u_2q)(\omega+u_3q)(\omega+u_4q)},
\end{equation}
\end{widetext}
where $u_{1,2,3,4}$ are the velocities of four independent branches
of the chiral excitations, and  they are all related to $g$, $\delta
v_R$ and $\delta v_B$.$^{14}$ Since linear response is exact for an
ideal LL, the external electric field has to be used for the
conductivity calculation,$^{18}$ i.e.,
\begin{equation}
\label{myeq19} j_\rho(q,\omega)=\sigma(q,\omega)E(q,\omega).
\end{equation}
Therefore, combining Eq. (\ref{myeq17}) with Eq. (\ref{myeq19}), we
obtain the nonlocal charge conductivity
\begin{widetext}
\begin{eqnarray}
\label{myeq20}
\sigma_\rho(q,\omega)=\frac{ie^2v_F}{\hbar\pi}\frac{2\omega(\omega^{2}-v_{\sigma}^{2}q^2)}
{(\omega^2-v_{\rho}^{2}q^2)(\omega^{2}-v_{\sigma}^{2}q^2)-(\delta
v_{R}q\omega-\frac{1}{2} \delta v_{B} v_{F}q^2-\frac{1}{2}
\frac{\delta v_{B}}{v_{F}}\omega^2)^2}.
\end{eqnarray}
\end{widetext}

On the other hand, the bosonic phase field $\vartheta_\sigma$ is
related to the spin current operator through
\begin{equation}
\label{myeq21}
j_\sigma=\sqrt{\frac{2}{\pi}}e\partial_t\vartheta_\sigma.
\end{equation}
Combing Eqs. (\ref{myeq15}) and ({\ref{myeq21}), the spin current
operator can be expressed as
\begin{widetext}
\begin{eqnarray}
\label{myeq22}
j_\sigma(q,\omega)=\frac{ie^2v_F}{\hbar\pi}\frac{2\omega(\delta
v_Rq\omega-\frac{1}{2} \delta v_Bv_Fq^2-\frac{1}{2} \frac{\delta
v_B}{v_F}\omega^2)
E(q,\omega)}{(\omega^2-v_{\rho}^{2}q^2)(\omega^{2}-v_{\sigma}^{2}q^2)-(\delta
v_{R}q\omega-\frac{1}{2} \delta v_{B} v_{F}q^2-\frac{1}{2}
\frac{\delta v_{B}}{v_{F}}\omega^2)^2}.
\end{eqnarray}
\end{widetext}
Therefore, using the linear response relation (\ref{myeq19}), we can
also obtain the spin conductivity \begin{widetext}
\begin{eqnarray}
\label{myeq23}
\sigma_\sigma(q,\omega)=\frac{ie^2v_F}{\hbar\pi}\frac{2\omega(\delta
v_Rq\omega-\frac{1}{2} \delta v_Bv_Fq^2-\frac{1}{2} \frac{\delta
v_B}{v_F}\omega^2)}{(\omega^2-v_{\rho}^{2}q^2)(\omega^{2}-v_{\sigma}^{2}q^2)-(\delta
v_{R}q\omega-\frac{1}{2} \delta v_{B} v_{F}q^2-\frac{1}{2}
\frac{\delta v_{B}}{v_{F}}\omega^2)^2}
\end{eqnarray}
\end{widetext}
which is also a function of $g$, $\alpha$, $B$, $\omega$ and $q$.

Next, for understanding the transport property of the system in more
detail, we go further for the two limited cases of $B=0$ or
$\alpha=0$, respectively, in the following subsections.

\subsection{The conductivity with Rashba SOC}
Consider a LL quantum wire submitted to Rashba SOC without a Zeeman
splitting, i.e., in the absence of external magnetic field ($B=0$ or
$\delta v_B=0$). In this case the expression of the current operator
Eq. (\ref{myeq17}) is reduced to
\begin{equation}
\label{myeq24} j_{\rho}(q,\omega)
=\frac{ie^2v_F}{\hbar\pi}\frac{2\omega(\omega^2-v_{\sigma}^{2}q^2)E(q,\omega)}
{(\omega^2-u_1^2q^2)(\omega^2-u_{2}^{2}q^2)},
\end{equation}
where
\begin{equation}
\label{myeq25} u_{1,2}^2=\frac{\delta
v_R^2+v_{\rho}^2+v_{\sigma}^2}{2}\pm \frac{\sqrt{(\delta
v_R^2+v_{\rho}^2+v_{\sigma}^2)^2-4v_{\rho}^2v_{\sigma}^2}}{2}
\end{equation}
are the propagation velocities of coupled collective modes in which
the subscript 1/2 corresponds to +/-. Furthermore, in the absence of
SOC ($\alpha=0$ or $\delta v_R=0$), we simply have
$u_{1,2}=v_{\rho,\sigma}$ which correspond to the velocities for the
special case of spin-charge separation in a LL quantum wire.$^{16}$
Moreover, Eq.(\ref{myeq24}) can be rewritten as
\begin{eqnarray}
\label{myeq26}
j_{\rho}(q,\omega)&=&\frac{ie^2v_{F}E(q,\omega)}{\hbar\pi}\Bigg[
\frac{u_{1}^2-v_{\sigma}^2}{u_{1}^2-u_{2}^2}\bigg(\frac{1}{\omega+u_1q}+
\frac{1}{\omega-u_1q}\bigg)\nonumber\\&-&\frac{u_{2}^2-v_{\sigma}^2}{u_{1}^2-u_{2}^2}\bigg(\frac{1}{\omega+u_2q}
+\frac{1}{\omega-u_2q}\bigg)\Bigg].
\end{eqnarray}

Therefore, combining Eq. (\ref{myeq19}) with Eq. (\ref{myeq26}), we
obtain the nonlocal charge conductivity
\begin{eqnarray}
\label{myeq27}
\sigma_\rho(q,\omega)&=&\frac{ie^2v_{F}}{\hbar\pi}\Bigg[
\frac{u_{1}^2-v_{\sigma}^2}{u_{1}^2-u_{2}^2}\bigg(\frac{1}{\omega+u_1q}+
\frac{1}{\omega-u_1q}\bigg)\nonumber\\&-&\frac{u_{2}^2-v_{\sigma}^2}{u_{1}^2-u_{2}^2}\bigg(\frac{1}{\omega+u_2q}
+\frac{1}{\omega-u_2q}\bigg)\Bigg],
\end{eqnarray}
which can be transformed into real space
\begin{equation}
\label{myeq28}
\sigma_\rho(x,\omega)=\frac{2e^2}{h}\Bigg[\frac{(u_{1}^2-v_{\sigma}^2)v_F}{(u_{1}^2-u_{2}^2)u_1}
e^{i\frac{\omega}{u_1}|x|}
-\frac{(u_{2}^2-v_{\sigma}^2)v_F}{(u_{1}^2-u_{2}^2)u_2}e^{i\frac{\omega}{u_2}|x|}\Bigg].
\end{equation}
For convenience we use the abbreviation $\xi=x/l$, in which $\xi$
provides a dimensionless measured position in the wire and $l$ is
the unit of length. Hence, Eq. (\ref{myeq28}) is reduced to
\begin{equation}
\label{myeq29}
\sigma_\rho(x,\omega)=\frac{2e^2}{h}\Bigg[\frac{(u_{1}^2-v_{\sigma}^2)v_F}{(u_{1}^2-u_{2}^2)u_1}e^{i\frac{
\omega
l}{u_1}|\xi|}-\frac{(u_{2}^2-v_{\sigma}^2)v_F}{(u_{1}^2-u_{2}^2)u_2}e^{i\frac{\omega
l}{u_2}|\xi|}\Bigg].
\end{equation}
This result implicits that the ac charge conductivity of a perfect
LL quantum wire with Rashba SOC is an oscillation function of the
interaction parameter, SOC strength and the driving frequency as
well as the measurement position in the wire. And in the limit of
vanishing spin-orbit and Zeeman coupling, the dc charge conductivity
$\sigma_\rho=2ge^2/h$ which is agreement with the conductivity of
the previous studies$^{18}$.

In addition, in the case of LL wire with Rashba SOC only, Eq.
(\ref{myeq22}) for spin current operator is reduced to
\begin{equation}
\label{myeq30}
j_\sigma(q,\omega)=\frac{ie^2v_F}{\hbar\pi}\frac{2\omega(-\delta
v_Rq\omega)E(q,\omega)}{(\omega^2-u_1^2q^2)(\omega^2-u_{2}^{2}q^2)},
\end{equation}
and similar calculation leads to the nonlocal spin conductivity for
the system
\begin{equation}
\label{myeq31} \sigma_\sigma(x,\omega)=\frac{2e^2}{h}\frac{\delta
v_Rv_F}{u_1^2-u_2^2}sign(\xi)\big(e^{i\frac{\omega
l}{u_1}|\xi|}-e^{i\frac{\omega l}{u_2}|\xi|}\big),
\end{equation}
where $sign(\xi)$= -1 for $\xi<0$ and 1 for $\xi>0$. This expression
for spin conductivity has a less complicated dependence on the
system parameters than charge conductivity (29). However, the spin
conductivity vanishes as $\delta v_R=0$ or $\omega=0$.

Furthermore, if we reverse the transformation (4) and define the
total (charge) conductivity
$\sigma_\rho=\sigma_\uparrow+\sigma_\downarrow$ and the difference
(spin) conductivity
$\sigma_\sigma=\sigma_\uparrow-\sigma_\downarrow$, then the
combination of Eqs. (\ref{myeq29}) and (\ref{myeq31}) gives
\begin{widetext}
\begin{eqnarray}
\label{myeq32}
\sigma_\uparrow&=&\frac{e^2}{h}\bigg[\Big(\frac{(u_{1}^2-v_{\sigma}^2)v_F}{(u_{1}^2-u_{2}^2)u_1}+\frac{\delta
v_Rv_F}{u_1^2-u_2^2}sign(\xi)\Big) e^{i\frac{\omega
l}{u_1}|\xi|}-\Big(\frac{(u_{2}^2-v_{\sigma}^2)v_F}{(u_{1}^2-u_{2}^2)u_2}+\frac{\delta
v_Rv_F}{u_1^2-u_2^2}sign(\xi)\Big)
e^{i\frac{\omega l}{u_2}|\xi|}\bigg]\nonumber\\
&\myequal{\omega\rightarrow0}&\frac{e^2}{h}\frac{v_F}{u_{1}^2-u_{2}^2}
\Big(\frac{u_{1}^2-v_{\sigma}^2}{u_{1}}-\frac{u_{2}^2-v_{\sigma}^2}{u_{2}}\Big)
\end{eqnarray}
and
\begin{eqnarray}
\label{myeq33}
\sigma_\downarrow&=&\frac{e^2}{h}\bigg[\Big(\frac{(u_{1}^2-v_{\sigma}^2)v_F}{(u_{1}^2-u_{2}^2)u_1}-\frac{\delta
v_Rv_F}{u_1^2-u_2^2}sign(\xi)\Big) e^{i\frac{\omega
l}{u_1}|\xi|}-\Big(\frac{(u_{2}^2-v_{\sigma}^2)v_F}{(u_{1}^2-u_{2}^2)u_2}-\frac{\delta
v_Rv_F}{u_1^2-u_2^2}sign(\xi)\Big)
e^{i\frac{\omega l}{u_2}|\xi|}\bigg]\nonumber\\
&\myequal{\omega\rightarrow0}&\frac{e^2}{h}\frac{v_F}{u_{1}^2-u_{2}^2}
\Big(\frac{u_{1}^2-v_{\sigma}^2}{u_{1}}-\frac{u_{2}^2-v_{\sigma}^2}{u_{2}}\Big)
\end{eqnarray}
\end{widetext}
for the conductivity of spin-up and spin-down electrons,
respectively. From Eqs. (32) and (33), we can see that in the case
of $\omega=0$ or without Rashba SOC ($\delta v_R$=0), the
conductivities for the two spin subband are degenerate. Defining
$v_{F\uparrow}=v_F-\delta v_R/2 ~(v_{F\downarrow}=v_F+\delta v_R/2)$
as the Fermi velocity of the spin-up(down) subband in the presence
of Rashba SOC, we can express $\delta v_R/v_F$ as
\begin{equation}
\label{myeq34} \frac{\delta v_R}{v_F}=\frac{2(\frac{\displaystyle
v_{F\downarrow}}{\displaystyle
v_{F\uparrow}}-1)}{\frac{\displaystyle
v_{F\downarrow}}{\displaystyle v_{F\uparrow}}+1},
\end{equation}
in which $v_{F\uparrow}=v_{F\downarrow}$ when $\delta v_R=0$.

\subsection{The conductivity with Zeeman splitting}
In this subsection we consider the case of the system with Zeeman
splitting in the absence of SOC, i.e., in the case of $\alpha=0$ or
$\delta v_R=0$. In this case the charge current operator reads
\begin{equation}
\label{myeq35} j_{\rho}(q,\omega)
=\frac{ie^2v_F}{\hbar\pi}\frac{2\omega(\omega^2-v_{\sigma}^{2}q^2)E(q,\omega)}
{\displaystyle [1-(\frac{\delta
v_B}{2v_{F}})^2](\omega^2-u_1^2q^2)(\omega^2-u_{2}^{2}q^2)},
\end{equation}
where
\begin{equation}
\label{myeq36} u_{1,2}^2=\frac{v_{\rho}^2+v_{\sigma}^2+\frac{\delta
v_B^2}{2}}{2[1-(\frac{\delta v_B}{2v_{F}})^2]}
\pm\sqrt{\bigg(\frac{v_{\rho}^2+v_{\sigma}^2+\frac{\delta v_B^2}{2}}
{2[1-(\frac{\delta v_B}{2v_{F}})^2]}\bigg)^2
-\frac{v_{\rho}^2v_{\sigma}^2-\frac{\delta
v_B^2v_{F}^2}{4}}{1-(\frac{\delta v_B}{2v_{F}})^2}}
\end{equation}
are the propagation velocities of collective modes. Again, when
$\delta v_B=0$, they are also reduced to the velocities for the
spin-charge separated system $u_{1,2}=v_{\rho,\sigma}$.$^{16,19}$

Additionally, through the same procedures as above, we obtain the
result for the nonlocal charge conductivity
\begin{eqnarray}
\label{myeq37}
\sigma_\rho(q,\omega)&=&\frac{ie^2v_F}{\hbar\pi}\frac{1}{1-(\frac{\delta
v_B}{2v_{F}})^2}\Bigg[
\frac{u_{1}^2-v_{\sigma}^2}{u_{1}^2-u_{2}^2}\bigg(
\frac{1}{\omega+u_1}+\frac{1}{\omega-u_1}\bigg)\nonumber\\
&-&\frac{u_{2}^2-v_{\sigma}^2}{u_{1}^2-u_{2}^2}\bigg(\frac{1}
{\omega+u_2}+\frac{1}{\omega-u_2}\bigg),
\end{eqnarray}
which can be transformed into real space
\begin{eqnarray}
\label{myeq38}
\sigma_\rho(x,\omega)&=&\frac{2e^2}{h}\frac{1}{1-(\frac{\delta
v_B}{2v_{F}})^2}
\Bigg[\frac{(u_{1}^2-v_{\sigma}^2)v_F}{(u_{1}^2-u_{2}^2)u_1}e^{i\frac{
\omega
l}{u_1}|\xi|}\nonumber\\&-&\frac{(u_{2}^2-v_{\sigma}^2)v_F}{(u_{1}^2-u_{2}^2)u_2}e^{i\frac{\omega
l}{u_2}|\xi|}\Bigg].
\end{eqnarray}
From this expression we can see that the ac conductivity of a
perfect LL with only Zeeman splitting is also an oscillation
function of the interaction parameter, magnetic field intensity and
the driving frequency as well as the measurement position in the
wire.

Accordingly, using the same method above the spin conductivity for
the LL wire only with Zeeman splitting is obtained as
\begin{eqnarray}
\label{myeq39}
\sigma_\sigma(x,\omega)&=&\frac{2e^2}{h}\frac{\frac{\delta
v_B}{2v_{F}}}{1-(\frac{\delta
v_B}{2v_{F}})^2}\Bigg[\frac{(u_{1}^2+v_{F}^2)v_F}{(u_{1}^2-u_{2}^2)u_1}e^{i\frac{
\omega
l}{u_1}|\xi|}\nonumber\\&-&\frac{(u_{2}^2+v_{F}^2)v_F}{(u_{1}^2-u_{2}^2)u_2}e^{i\frac{\omega
l}{u_2}|\xi|}\Bigg].
\end{eqnarray}
We also see that as $\delta v_B=0$ the spin conductivity vanishes.
Under the same definition of the total (charge) conductivity
$\sigma_\rho=\sigma_\uparrow+\sigma_\downarrow$ and the difference
(spin) conductivity
$\sigma_\sigma=\sigma_\uparrow-\sigma_\downarrow$ with the
combination of Eqs. (\ref{myeq38}) and (\ref{myeq39}), we can obtain
\begin{widetext}
\begin{eqnarray}
\label{myeq40}
\sigma_\uparrow&=&\frac{e^2}{h}\frac{1}{1-(\frac{\delta
v_B}{2v_{F}})^2}\frac{v_F}{u_{1}^2-u_{2}^2}\Bigg[\Big(\frac{u_{1}^2-v_{\sigma}^2}{u_1}
+\frac{\delta v_B}{2v_F}\frac{u_{1}^2+v_{F}^2}{u_1}\Big)e^{i\frac{
\omega l}{u_1}|\xi|}-\Big(\frac{u_{2}^2-v_{\sigma}^2}{u_2}
+\frac{\delta v_B}{2v_F}\frac{u_{2}^2+v_{F}^2}{u_2}\Big)e^{i\frac{
\omega
l}{u_2}|\xi|}\Bigg]\nonumber\\
&\myequal{\omega\rightarrow0}&\frac{e^2}{h}\frac{1}{1-(\frac{\delta
v_B}{2v_{F}})^2}\frac{v_F}{u_{1}^2-u_{2}^2}\Bigg[\Big(\frac{u_{1}^2-v_{\sigma}^2}{u_1}
+\frac{\delta
v_B}{2v_F}\frac{u_{1}^2+v_{F}^2}{u_1}\Big)-\Big(\frac{u_{2}^2-v_{\sigma}^2}{u_2}
+\frac{\delta v_B}{2v_F}\frac{u_{2}^2+v_{F}^2}{u_2}\Big)\Bigg]
\end{eqnarray}
and
\begin{eqnarray}
\label{myeq41}
\sigma_\downarrow&=&\frac{e^2}{h}\frac{1}{1-(\frac{\delta
v_B}{2v_{F}})^2}\frac{v_F}{u_{1}^2-u_{2}^2}\Bigg[\Big(\frac{u_{1}^2-v_{\sigma}^2}{u_1}
-\frac{\delta v_B}{2v_F}\frac{u_{1}^2+v_{F}^2}{u_1}\Big)e^{i\frac{
\omega l}{u_1}|\xi|}-\Big(\frac{u_{2}^2-v_{\sigma}^2}{u_2}
-\frac{\delta v_B}{2v_F}\frac{u_{2}^2+v_{F}^2}{u_2}\Big)e^{i\frac{
\omega
l}{u_2}|\xi|}\Bigg]\nonumber\\
&\myequal{\omega\rightarrow0}&\frac{e^2}{h}\frac{1}{1-(\frac{\delta
v_B}{2v_{F}})^2}\frac{v_F}{u_{1}^2-u_{2}^2}\Bigg[\Big(\frac{u_{1}^2-v_{\sigma}^2}{u_1}
-\frac{\delta
v_B}{2v_F}\frac{u_{1}^2+v_{F}^2}{u_1}\Big)-\Big(\frac{u_{2}^2-v_{\sigma}^2}{u_2}
-\frac{\delta v_B}{2v_F}\frac{u_{2}^2+v_{F}^2}{u_2}\Big)\Bigg]
\end{eqnarray}
\end{widetext}
for the conductivities of spin-up and spin-down electrons,
respectively. We also see that without Zeeman splitting ($\delta
v_B$=0) the conductivities for the two spin subband are degenerate.
Again, defining $v_{F\uparrow}=v_F+\delta v_B/2
~(v_{F\downarrow}=v_F-\delta v_B/2)$ as the Fermi velocity of the up
(down) spin subband in a magnetic field $B$, we can also express
$\delta v_B/v_F$ as
\begin{equation}
\label{myeq42} \frac{\delta v_B}{v_F}=\frac{2(\frac{\displaystyle
v_{F\uparrow}}{\displaystyle
v_{F\downarrow}}-1)}{\frac{\displaystyle
v_{F\uparrow}}{\displaystyle v_{F\downarrow}}+1},
\end{equation}
which has the similar form as Eq. (34).

\section{Results and discussions}
There are six calculated figures presented in this paper, in which
Fig. 1 to Fig. 3 are plotted for the system with Rashba SOC in the
absence of an external magnetic field, whereas Fig. 4 to Fig. 6 are
plotted for the system with Zeeman splitting in the absence of
Rashba SOC.

\begin{figure}
\center
\includegraphics[width=2.3in]{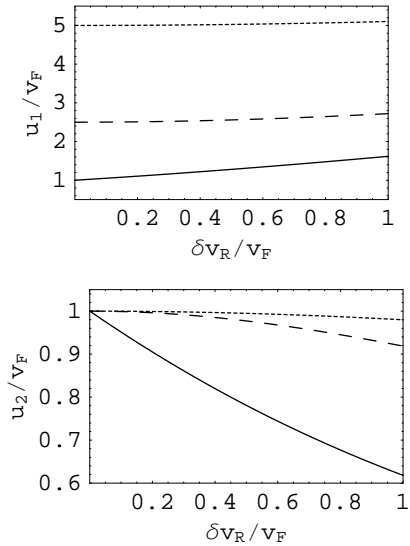}
\caption{The plotted propagation velocities of the collective modes
$u_{1,2}$ (in units of $v_F$) as a function of the SOC strength
$\delta v_R/v_F$ in the absence of Zeeman effect, where the solid
lines for $g=1$, the dashed lines for $g=0.4$ and the dotted lines
for $g=0.2$, respectively.}
\end{figure}

The dimensionless velocities of the bosonic excitation $u_1 (u_2)$
(in units of $v_F$) as a function of $\delta v_R/v_F$ calculated
according to Eq. (25) in the absence of Zeeman splitting for three
different electron-electron interaction strengths of $g$=0.2 (dotted
line), 0.4 (dashed line) and 1 (solid line), respectively, are shown
in Fig. 1. We can see that when the interaction is turned on ($g<1$)
and as $\delta v_R/v_F$ (proportional to Rashba SOC strength)
increases $u_1/v_F$ increases while $u_2/v_F$ decreases slightly.
However, for the stronger interaction the changes of $u_1/v_F$ and
$u_2/v_F$ seem less obvious. For a fixed value of $\delta v_R/v_F$,
the stronger the interaction is, the larger the velocities of the
bosonic excitation $u_1 (u_2)$ are. And in the limited case of
$\delta v_R/v_F=0$ (i.e., in the absence of SOC), $u_1/v_F$ is equal
to $1/g$ whereas $u_2/v_F$ is equal to $1$. This is the known result
that can be found in Ref. 20.

\begin{figure}
\center
\includegraphics[width=2.3in]{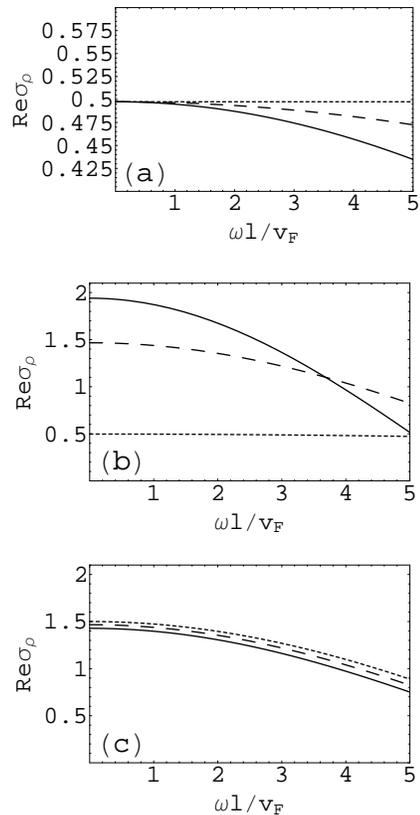}
\caption{The plotted $Re\sigma_\rho(x,\omega)$ (in units of $e^2/h$)
as a function of $\omega l/v_F$ in the absence of Zeeman effect (a)
with fixed $g=0.25$ and $\delta v_R/v_F=0.5$ where the solid line
for $\xi=\pm0.4$, the dashed line for $\xi=\pm0.25$ and the dotted
line for $\xi=0$; (b) with fixed $\xi=0.25$ and $\delta v_R/v_F=0.5$
where the solid line for $g=1$, the dashed line for $g=0.75$ and the
dotted line for $g=0.25$; and (c) with fixed $\xi=0.25$ and $g=0.75$
where the solid line for $\delta v_R/v_F=0.75$, the dashed line for
$\delta v_R/v_F=0.5$ and the dotted line for $\delta v_R/v_F=0$,
respectively.}
\end{figure}

Figure 2 illustrates the real part of charge conductivity
$Re\sigma_\rho(x,\omega)$ (in units of $e^2/h$) as a function of
$\omega l/v_F$ calculated according to Eq. (29) in the absence of
Zeeman splitting. For the system with fixed $g=0.25$ and $\delta
v_R/v_F=0.5$, Fig. 2(a) shows the dependence of three different
measurement positions $\xi=\pm0.4$ (solid line), $\pm0.25$ (dashed
line) and 0 (dotted line) on the conductivity, respectively. In the
center of the wire $(\xi=0)$ the conductivity is a constant value
regardless of $\omega l/v_F$. However, the further off the wire
center the position, the quicker the change of the conductivity is.
Notice that the conductivity only depends on the absolute value
$|\xi|$. The influence of the electron-electron interaction $g$ and
the Rashba strength $\delta v_R/v_F$ on $Re\sigma_\rho(x,\omega)$
for the system are shown in Fig. 2(b) and Fig. 2(c), respectively.
For the system with fixed $\xi=0.25$ and $\delta v_R/v_F=0.5$, Fig.
2(b) shows the dependence of three different interaction strengths
$g=1$ (solid line), 0.75 (dashed line) and 0.25 (dotted line) on the
conductivity, respectively. When the electron-electron interaction
parameter $g$ is larger, the variation of $Re\sigma_\rho(x,\omega)$
is faster. If the compositive vibration has a periodicity, then the
stronger the interaction is, the longer the period of the
oscillation is. The period of the oscillation is the least common
multiple of $2\pi u_1/(v_F |\xi|)$ and $2\pi u_2/(v_F |\xi|)$, and a
measured position is fixed $\xi=0.25$, so the period of the
oscillation is totally determined by $u_1$ and $u_2$. For the system
with fixed $g=0.75$ and $\xi=0.25$, Fig. 2(c) shows the dependence
of three different Rashba strengths $\delta v_R/v_F=0.75$ (solid
line), 0.5 (dashed line) and 0 (dotted line) on the conductivity,
respectively. Comparing Fig. 2(c) with 2(b), we can find that the
dependence of $Re\sigma_\rho(x,\omega)$ on $g$ at fixed $\delta
v_R/v_F$ is very similar to that on $\delta v_R/v_F$ at fixed $g$,
and they exhibit the same tendency as a function of $\omega l/v_F$.
But it is obvious that the modification due to the electron-electron
interactions is remarkable. Moreover, from Figs. 2(b) and 2(c), we
can see that the dc ($\omega=0$) charge conductivities of the system
with different electron-electron interactions and Rashba strengths
are different constant values, which can be obtained analytically
from Eq.(\ref{myeq29}) with
$\lim_{\omega\rightarrow0}\sigma_\rho(x,\omega)=
2e^2/h[(u_1^2-v_\sigma^2)v_F/u_1/(u_1^2-u_2^2)-(u_2^2-v_\sigma^2)v_F/u_2/(u_1^2-u_2^2)]$.
Here $u_1$ and $u_2$ are dependent on both $g$ and $\alpha$ (see
Fig. 1).

\begin{figure}
\center
\includegraphics[width=2.3in]{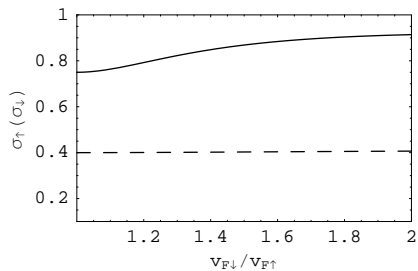}
\caption{The plotted spin-polarized charge conductivity
$\sigma_\uparrow(\sigma_\downarrow)$ (in units of $e^2/h$) as a
function of the ratio $v_{F\downarrow}/v_{F\uparrow}$ in the absence
of Zeeman effect, where the solid lines correspond to $g=0.75$ and
the dashed lines to $g=0.4$, respectively.}
\end{figure}

In Fig. 3 we show the dependence of spin-polarized dc conductivity
$\sigma_\uparrow (\sigma_\downarrow)$ on the ratio
$v_{F\downarrow}/v_{F\uparrow}$ (also proportional to Rashba SOC
strength) in the absence of Zeeman splitting for the two different
interaction strengths, where the solid lines correspond to $g=0.75$
and the dashed lines to $g=0.4$, respectively. The curves for
$\sigma_\downarrow$ as the function of
$v_{F\downarrow}/v_{F\uparrow}$ are the same as those for
$\sigma_\uparrow$, and this can be verified from Eqs. (\ref{myeq32})
and (\ref{myeq33}). This result implies that the dc conductivities
of spin-up and spin-down electrons are degenerate in the absence of
Zeeman effect. From Fig. 3 we can also see that the spin-polarized
conductivities $\sigma_\uparrow (\sigma_\downarrow)$ is connected
with the electron-electron interactions for any fixed value of
$v_{F\downarrow}/v_{F\uparrow}$. The stronger the interactions
produce the smaller $\sigma_\uparrow$ ($\sigma_\downarrow$) which
show that the repulsive interaction suppresses the conductivity. And
the less obvious the increase of $\sigma_\uparrow$
($\sigma_\downarrow$) as the increases of ratio
$v_{F\downarrow}/v_{F\uparrow}$, which show that in the case of
strong electron-electron interaction the modification of
$v_{F\downarrow}/v_{F\uparrow}$ on the conductivity is very little.

\begin{figure}
\center
\includegraphics[width=2.3in]{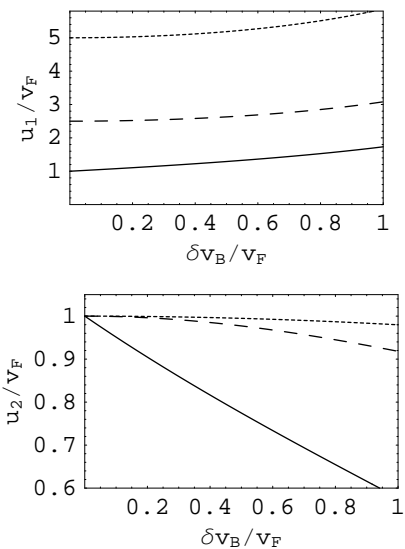}
\caption{The plotted propagation velocities of the collective modes
$u_{1,2}$ (in units of $v_F$) as a function of the Zeeman strength
$\delta v_B/v_F$ in the absence of SOC, where the solid lines
correspond to $g=1$, the dashed lines to $g=0.4$ and the dotted
lines to $g=0.2$, respectively.}
\end{figure}

Figure 4 shows the dimensionless velocities of the bosonic
excitations $u_1(u_2)/v_F$ vs $\delta v_B/v_F$ of the system
calculated according to Eq. (36) in the absence of Rashba SOC for
three different interaction strengths of $g$=1 (solid line), 0.4
(dashed line) and 0.2 (dotted line). We also see that with the
increase of $\delta v_B/v_F$  $u_1/v_F$ increases while $u_2/v_F$
decreases slightly for all values of $g<1$. But for noninteracting
case ($g$=1) $u_2/v_F$ decays more rapidly as $\delta v_B/v_F$
increases, and the stronger the interaction is, the slower the decay
of $u_2/v_F$. Fig. 4 is very similar to Fig. 1, which makes clear
that the virtual magnetic field induced by an electric field
perpendicular to the 2DEG yields the similar effect on the
propagation velocities of the collective modes as the magnetic field
applied along the $y$ direction.

\begin{figure}
\center
\includegraphics[width=2.3in]{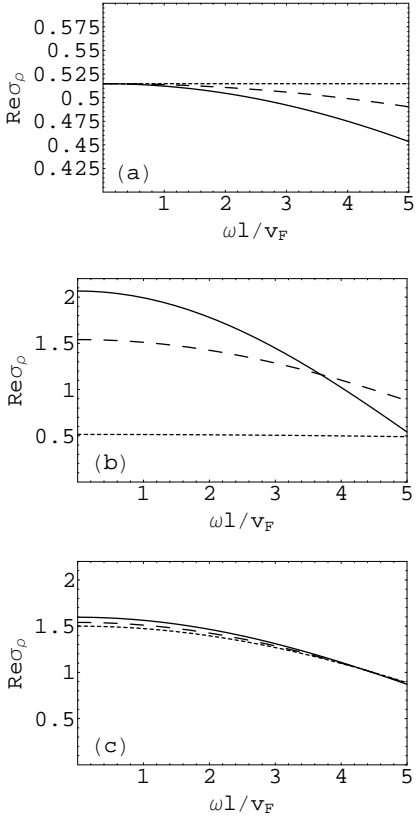}
\caption{The plotted $Re\sigma_\rho(x,\omega)$ (in units of $e^2/h$)
as a function of $\omega l/v_F$ in the absence of SOC (a) with fixed
$g=0.25$ and $\delta v_B/v_F=0.5$ where the solid line for
$\xi=\pm0.4$, the dashed line for $\xi=\pm0.25$ and the dotted line
for $\xi=0$; (b) with fixed $\xi=0.25$ and $\delta v_B/v_F=0.5$
where the solid line for $g=1$, the dashed line for $g=0.75$ and the
dotted line for $g=0.25$; and (c) with fixed $\xi=0.25$ and $g=0.75$
where the solid line for $\delta v_B/v_F=0.75$, the dashed line for
$\delta v_B/v_F=0.5$ and the dotted line for $\delta v_B/v_F=0$,
respectively.}
\end{figure}

Figure 5 illustrates the real part of charge conductivity
$Re\sigma_\rho(x,\omega)$ (in units of $e^2/h$) as a function of
$\omega l/v_F$ calculated according to Eq. (38) in the absence of
Rashba SOC. For the system with fixed $g=0.25$ and $\delta
v_B/v_F=0.5$, Fig. 5(a) shows the dependence of three different
measurement positions $\xi=\pm0.4$ (solid line), $\pm0.25$ (dashed
line) and 0 (dotted line) on the conductivity, respectively. Figure
5(b) shows the dependence of three different interaction strengths
$g=1$ (solid line), 0.75 (dashed line) and 0.25 (dotted line) on the
conductivity for the system with fixed $\xi=0.25$ and $\delta
v_B/v_F=0.5$. And for the system with fixed $g=0.75$ and $\xi=0.25$,
Fig. 5(c) shows the dependence of three different magnetic strengths
$\delta v_B/v_F=0.75$ (solid line), 0.5 (dashed line) and 0 (dotted
line) on the conductivity, respectively. The variables and the
scales in Fig. 5 are the same as that in Fig. 2 except replacing
$\delta v_R/v_F$ by $\delta v_B/v_F$. And Fig. 5 is very similar to
Fig. 2. However, from Fig. 5 we can see that the dc ($\omega=0$)
charge conductivities of the system with different electron-electron
interactions or magnetic strengths are different constant values
[Fig. 5(b) and 5(c)], but with different measurement positions are
the same values [Fig. 5(a)]. This is because that
$\lim_{\omega\rightarrow0}\sigma_\rho(x,\omega)=2e^2/h(1/[1-(\delta
v_B/(2v_F))^2])[(u_1^2-v_\sigma^2)v_F/u_1/(u_1^2-u_2^2)-(u_2^2-v_\sigma^2)v_F/u_2/(u_1^2-u_2^2)]$
where $u_1$ and $u_2$ are only dependent on both $g$ and $\delta
v_B$ (see Fig. 4). These constant values are almost the product of
$Re\sigma_\rho(x,\omega=0)$ in Fig. 2 and a factor $1/[1-(\delta
v_B/2v_F)^2]$, and is also true in the case of $\omega\neq0$.
Comparing Fig. 5(c) with 5(b), we can find that the dependence of
$Re\sigma_\rho(x,\omega)$ on $g$ at fixed $\delta v_B/v_F$ is very
similar to that on $\delta v_B/v_F$ at fixed $g$, and they exhibit
the same tendency as a function of $\omega l/v_F$. But it is obvious
that the modification due to the electron-electron interactions is
more remarkable, which is the same as the conclusion by comparing
Fig. 2(c) with 2(b). Moreover comparing Fig. 5(c) with 2(c), we can
find that the dependence of $Re\sigma_\rho(x,\omega)$ on $\delta
v_B/v_F$ is very similar to that on $\delta v_R/v_F$ in the case of
fixed $g$, and they exhibit the same tendency as a function of
$\omega l/v_F$. This means that the effects of Rashba SOC and Zeeman
splitting on the conductivity are very similar.
\begin{figure}
\center
\includegraphics[width=2.3in]{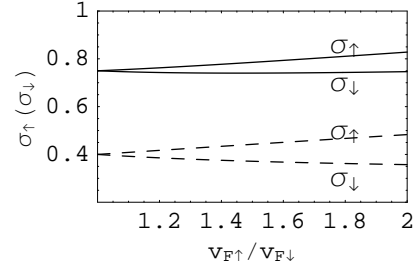}
\caption{The plotted spin-polarized charge conductivity
$\sigma_\uparrow(\sigma_\downarrow)$ (in units of $e^2/h$) as a
function of the ratio $v_{F\uparrow}/v_{F\downarrow}$ in the absence
of SOC, where the solid lines correspond to $g=0.75$ and the dashed
lines to $g=0.4$, respectively.}
\end{figure}

In Fig. 6 we show how the spin polarized dc conductivities
$\sigma_\uparrow$ and $\sigma_\downarrow$ evolve as the ratio of
$v_{F\uparrow}/v_{F\downarrow}$ is varied for the two different
interaction strengths, where the solid lines correspond to $g=0.75$
and the dashed lines to $g=0.4$, respectively. It is also
demonstrated that the increase of ratio
$v_{F\uparrow}/v_{F\downarrow}$ pushes $\sigma_\uparrow$ and
$\sigma_\downarrow$ away from each other, and one accelerates while
the other slows down. This result is in agreement with Ref. 9. In
contrast to the Rashba SOC case as shown in Fig. 3, the ratio of the
spin-polarized conductivities $\sigma_\uparrow/\sigma_\downarrow$ is
dependent on the electron-electron interactions and the ratio
$v_{F\uparrow}/v_{F\downarrow}$. This is because the channel with a
larger electron velocity has a larger transmission coefficient for
the fixed electron-electron interaction strength and
$v_{F\downarrow}/v_{F\uparrow}$, and the difference of the
transmission coefficient between channels becomes larger with the
increase of $v_{F\downarrow}/v_{F\uparrow}$ or of electron-electron
interaction strength.

Finally, in II-VI semiconductors the Rashba SOC is expected to be
larger than the Dresselhaus coupling, so we can neglect the
Dresselhaus SOC.$^8$ At low temperatures, 2DEG formed in II-VI
semiconductor heterostructures is restricted by a transverse
confining potential, so we have the narrow long enough quantum wire.
A weak magnetic field $B$ perpendicular to the quantum wire is
applied along $y$-axis, which is turned on or turned off according
to the requirement. A longitudinally polarized external
electromagnetic field with wavevector along the $z$-axis irradiates
the quantum wire. For the aforementioned conditions, the present
experiment can be available.$^{2-4}$ Furthermore, in this article we
have only considered the case of infinite-length LL.  The situation
is important for a simple theoretical understanding, although it may
be relevant to experiments where the leads always dominate the
results. The conductivity of a finite-size LL coupled to leads may
have different frequency and amplitude dependence on the physical
parameters of the system due to the Andreev-type reflections.$^{19}$
The detailed calculation and discussion for the important role the
Fermi liquid leads on the conductivity of the system will be given
in our next work.

\section{Conclusion}
In conclusion, using a straightforward approach we have investigated
theoretically the transport properties through an interacting
quantum wire in the presence of both Rashba SOC and Zeeman splitting
simultaneously in the LL regime. Using the bosonization technique,
the motion of equations of bosonic phase fields for the system with
a longitudinal electric field is established, and its solution is
obtained by introducing a Fourier transformation in which the spin
and charge degrees of freedom are completely coupled and
characterized by four new different velocities. Within the linear
response theory, it is found that the ac conductivity of a LL wire
in the presence of Rashba SOC and Zeeman splitting is generally an
oscillation function of the interaction strength $g$, Rashba SOC
strength $\alpha$, Zeeman interaction strength $B$ and the driving
frequency $\omega$ as well as the measurement position $x$ in the
wire.

For a LL wire only with Rashba SOC, the real part of the
conductivity $Re\sigma_\rho$ as the function of the
electron-electron interaction or Rashba SOC strength exhibits the
similar decay tendencies with the increase of frequency $\omega
l/v_F$. But the modification due to the electron-electron
interactions is more remarkable than that due to Rashba SOC. On the
other hand, for a LL wire only with Zeeman splitting. On the other
hand, for a LL wire only with Zeeman splitting, $Re\sigma_\rho$ is a
function of the frequency $\omega l/v_F$, the same curves hold if
one replaces $\delta v_R/v_F$ by $\delta v_B/v_F$, but one has to
multiply all values of $Re\sigma_\rho$ by a factor of $1/[1-(\delta
v_B/2v_F)^2]$.

For a LL only with Rashba SOC, when we study how $\sigma_\uparrow
(\sigma_\downarrow)$ evolve as the ratio
$v_{F\downarrow}/v_{F\uparrow}$ is varied, we find that the curves
for $\sigma_\downarrow$ as the function of
$v_{F\downarrow}/v_{F\uparrow}$ are the same as those for
$\sigma_\uparrow$. But for a LL only with Zeeman splitting, we find
that the increase of the ratio $v_{F\uparrow}/v_{F\downarrow}$
pushes $\sigma_\uparrow$ and $\sigma_\downarrow$ away from each
other, which is consistent with that of the previous studies$^9$ for
the same system. In contrast to the SOC case, the ratio of the
spin-polarized conductivities $\sigma_\uparrow/\sigma_\downarrow$ is
dependent on the electron-electron interactions.

Further investigations are worthy to be done for the higher
conductivity corrections, in the presence of impurity, or with
realistic Coulomb interactions.

\begin{acknowledgements}
This work was supported by National Natural Science Foundation of
China (Grant No. 10574042), Specialized Research Fund for the
Doctoral Program of Higher Education of China (Grant No.
20060542002) and Hunan Provincial Natural Science Foundation of
China (Grant No. 06JJ2097).
\end{acknowledgements}

%\newpage

\end{document}